\documentclass[conference, 10pt]{IEEEtran}

\usepackage{cite}
\usepackage{url}
\usepackage{graphicx}
\usepackage{color}
\usepackage{placeins}
\usepackage{float}
\usepackage{tabularx,colortbl}
\usepackage{ifthen}
\usepackage{hhline}
\usepackage{lipsum}

\newcommand\blfootnote[1]{%
  \begingroup
  \renewcommand\thefootnote{}\footnote{#1}%
  \addtocounter{footnote}{-1}%
  \endgroup
}
\makeatletter

\newcounter{author}
\renewcommand{\author}[2][]{
   \stepcounter{author}
   \@namedef{author@\theauthor}{#2}
   \@namedef{authorlabel@\theauthor}{#1}
}

\newcounter{address}
\newcommand{\address}[2][]{
   \stepcounter{address}
   \@namedef{address@\theaddress}{#2}
   \@namedef{addresslabel@\theaddress}{#1}
}

\newcommand{\alsep}{and}

\def\newmaketitle{\par%
  \begingroup%
  \normalfont%
  \def\thefootnote{}
  \def\footnotemark{}
  \let\@makefnmark\relax
  \footnotesize
  \footnotesep 0.7\baselineskip
  \normalsize%
  \twocolumn[\thenewmaketitle\@IEEEaftertitletext]%
  \if@IEEEusingpubid
     \enlargethispage{-\@IEEEpubidpullup}%
  \fi
  \endgroup
  \setcounter{footnote}{0}\let\maketitle\relax\let\@maketitle\relax
  \gdef\@thanks{}%
  \let\thanks\relax}

\def\thenewmaketitle{
  \newpage
  \begin{center}%
    \vskip0.2em{\Huge\@IEEEcompsoconly{\sffamily}\@IEEEcompsocconfonly{\normalfont\normalsize\vskip 2\@IEEEnormalsizeunitybaselineskip
   \bfseries\large}\@title\par}\vskip1.0em\par%
    \vspace{1ex}
    \newcounter{c@author}
    \newcounter{c@tmp}
    \ifthenelse{\value{author}=2}{%
      \newcommand{\liand}{ and }}{%
      \newcommand{\liand}{, and }}
    \ifthenelse{\value{address}<2}{%
      \@nameuse{author@1}%
      \stepcounter{c@author}%
      \whiledo{\value{c@author}<\value{author}}{%
        \setcounter{c@tmp}{\value{author}}%
        \addtocounter{c@tmp}{-\value{c@author}}%
        \ifthenelse{\value{c@tmp}=1}{%
          \renewcommand{\alsep}{\liand}}{\renewcommand{\alsep}{, }}%
        \stepcounter{c@author}\alsep \@nameuse{author@\thec@author}}\\%
    }
    {
      \@nameuse{author@1}${}^{(\ref{\@nameuse{authorlabel@1}})}$%
      \stepcounter{c@author}%
      \whiledo{\value{c@author}<\value{author}}{%
      \setcounter{c@tmp}{\value{author}}%
      \addtocounter{c@tmp}{-\value{c@author}}%
      \ifthenelse{\value{c@tmp}=1}{%
        \renewcommand{\alsep}{\liand}}{\renewcommand{\alsep}{, }}%
      \stepcounter{c@author}\alsep \@nameuse{author@\thec@author}%
        ${}^{(\ref{\@nameuse{authorlabel@\thec@author}})}$%
      }
    }
    \vspace{0.2ex}

    \ifthenelse{\value{address}>0}{%
      \ifthenelse{\value{address}=1}{
        {\@nameuse{address@1}}
      }
      {
        \newcounter{c@address}

        \begin{center}
        \whiledo{\value{c@address}<\value{address}}
        {
          \refstepcounter{c@address}
            ${}^{(\thec@address)}$\,%
              \label{\@nameuse{addresslabel@\thec@address}}%
              \@nameuse{address@\thec@address}\\ %
        }
        \end{center}
      } 
    }
    {
      \relax
    }
  \end{center}
}

\makeatother

\thanks{Manuscript received November 22, 2021; accepted December 19, 2022. This work was supported in part by the National Science Foundation under award CNS-1439682-011 (BWAC IUCRC) and the Department of the Navy, Office of Naval Research under contract N00014-20-C-1067. Distribution A: Approved for public release: Distribution Unlimited. \textit{Corresponding author: Jonathan D. Chisum.}}\thanks{Wei Wang and Nicholas Estes contributed equally as lead authors.}\thanks{Wei Wang, Nicholas Estes, Nicolas Garcia, and Jonathan Chisum are with the Department of Electrical Engineering, University of Notre Dame, Notre Dame, IN, 46556 USA (e-mail: wwang23@nd.edu; nestes@nd.edu; abolstad@iastate.edu; jchisum@nd.edu). Matthew Roddy was with the Department of Electrical Engineering, University of Notre Dame at the time of the work and is now with Cornell University, Ithaca, NY, 14850 USA. Andrew Bolstad is with the Department of Electrical and Computer Engineering, Iowa State University, Ames, IA, 50011, USA.}

\title{High-frequency Limits for 3D-Printed Gradient-index (GRIN) Lens Antennas}

\author[org1]{Wei Wang}
\author[org2]{Philip Lambert}
\author[org1]{Jonathan Chisum}

\address[org1]{The University of Notre Dame, IN 46556, USA, (wwang23@nd.edu, jchisum@nd.edu)}
\address[org2]{3D Fortify, Inc, Boston, MA 02129, USA, (phillambert@3dfortify.com)}

\begin{document}

\newmaketitle

\begin{abstract}
Artificial dielectrics are widely used for Gradient-Index (GRIN) lens antennas. The unit-cell size of an artificial dielectric determines the maximum operating frequency and also drives cost and yield. To explore the frequency limitations we printed four identical Luneburg lens antennas using gyroid unit-cells of 12.5, 10, 7.5, and 5\,mm and measured their gain over the K- and Ka-band. We find maximum frequencies of 20, 25, 33, and $>$40\,GHz for each unit-cell, respectively. These measurements suggest a print resolution limit of $0.7\lambda_g$, where $\lambda_g$ is the wavelength in the host dielectric. 
\end{abstract}

\section{Introduction}
\begin{figure}[t]

\centering
  \includegraphics[width=0.9\columnwidth]{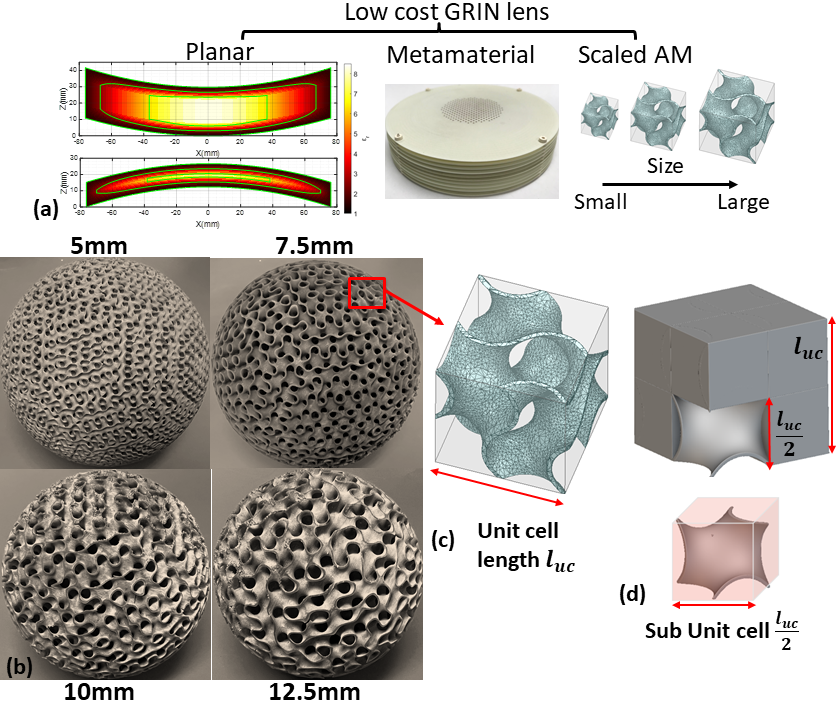}
  \caption{(a) Low-cost GRIN lens concepts: planar\cite{Garcia2022CompundPPWG}, metamaterial \cite{AnastasioGRIN} and scaled AM. (b) 3D-printed Luneburg lens photos. (c) The cubic gyroid structure unit-cell. (d) Ihe decomposed sub-unit cell of the gyroid.}\label{Fig1Label}
\vspace{-0.5cm}
\label{Fig:concept}
\end{figure}

Gradient-Index (GRIN) lens antennas \blfootnote{This work was supported in part by the Office of Naval Research under Contract \#N00014-20-C-1067. DISTRIBUTION STATEMENT A. Approved for Public Release.}offer a low-power and low-cost solution for high gain millimeter-wave (MMW) antennas. While low-power and high-gain are intrinsic to all lens antennas, low-cost is only possible if a simple manufacturing method can be realized. Figure\,\ref{Fig:concept} depicts various low-cost GRIN lens solutions including low-profile lenses and architectures like the compound, high-index lenses\cite{Garcia2022CompundPPWG,Xu2021HalfLLlens} which reduce overall lens volume to reduce cost, or metamaterial-based lenses which utilize standard PCB-based manufacturing to realize a gradient index\cite{AnastasioGRIN}\cite{Su2018metalens}. Additive manufacturing (AM) can also be used for low-cost manufacturing. However, since AM print resolution is limited ,this paper investigates the high-frequency performance of AM GRIN lenses for various unit-cell sizes to determine the maximum frequency of operation of a GRIN lens for a given unit-cell size.


For example, the perforated dielectric is fabricated by drilling holes in a given host material. A larger unit-cell size results in less drilling process in the fabrication\cite{Garcia2022CompundPPWG}. In \cite{Garcia_Optimizer_2021} the perforated unit-cell is relaxed from $\lambda_g/8$ to $\lambda_g/5$ without compromising aperture efficiency. This increase in unit-cell size represents a cost savings of 2.5$\times$ for a 2D lens. However, perforated dielectrics require a serial manufacturing process (i.e., drilling) and will therefore never provide an extremely low-cost solution to GRIN lens antenna manufacturing. On the other hand, recent advances in parallel AM with low-loss dielectric resins offer a key enabling technology for low-cost and high-frequency lens antennas. For example, the Fortify 3D-printer with the Rogers Radix(TM) dielectric resin offers low dielectric loss with the ability to print an entire lens layer in a single exposure\cite{Lambert2022Fortify}. 

In general, larger unit-cells have higher print yield and offer the potential for an increased range of effective permittivities so it is important to characterize the maximum frequency of operation for a given unit-cell size. As operating wavelength becomes comparable to the unit-cell size, the effective permittivity approximation breaks down causing reduced antenna gain. Therefore, choosing a proper unit-cell size for 3D printing is crucial to achieving an optimal cost-performance ratio at high frequencies. Various unit-cell sizing criterion are used in the literature including $0.1\lambda_o$\cite{Erfani2016_0p1}, $0.2\lambda_o$\cite{Xu2021_0p2UC} and $0.43\lambda_o$\cite{Jiang2017_0p43UC}, but these are more of a practical choice, and a systematic unit-cell size investigation is valuable for the low-cost 3D printing application. This paper presents the performance limits of 3D-printed GRIN lenses versus unit-cell size for a classic Luneburg lens model (chosen for generality). Section\,\ref{sec:lensProfile} discusses the Luneburg lens profile, the unit-cell structure and the lens fabrication. Section\,\ref{sec:meas} reports gain performance of four printed Luneburg lenses with various unit-cell sizes, and their performance limits are analyzed.

\section{Luneburg Lens Design Prototypes} \label{sec:lensProfile}


The classic Luneburg lens permittivity profile is a continuous function of radius, $\epsilon(r) = 2 - \left(r/R\right)^2$, where $R$ is the radius of the Luneburg lens and $r$ is the distance from the lens center. To realize the Luneburg profile a gyroid unit-cell is used on a cubic lattice with length $l_{uc}$ as shown in Fig\,\ref{Fig:concept}(c). The full-size gyroid can be further decomposed into a $l_{uc}/2$ sub-unit cell, shown in Fig\,\ref{Fig:concept}(d).
The effective permittivity of the gyroid unit-cell is varied by changing wall thickness\cite{Lambert2022Fortify}.




\begin{figure}[t]

\centering
  \includegraphics[width=1.05\columnwidth]{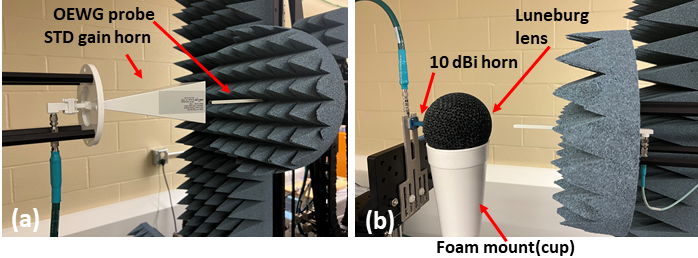}
  \caption{(a) Standard gain horn measurement. (b) Luneburg lens measurement.}\label{Fig1Label}
\vspace*{-0.15in}
\label{Fig:setup}
\end{figure} 


Four identical Luneburg lenses are fabricated using the gyroid structure with 10\,cm diameter. The cubic unit-cell for each of the four prototypes have $l_{uc}$ of 5, 7.5, 10, and 12.5\,mm, respectively. The lenses are printed using a Fortify Flux Core DLP 3D-printer. The material used in the measurement is Rogers Corporation Radix(TM) 3D-printable polymer dielectric with a relative permittivity of 2.8. In order to maintain structural integrity of the unit-cell, a minimum sidewall thickness limits the minimum effective permittivity to 1.2. Therefore the Luneburg lens permittivity profile was printed such that all radii for which nominal permittivity is less than 1.2 would maintain that value to the periphery. The printed Luneburg lenses are shown in Fig.\,\ref{Fig:concept}(b). 

\section{Measurement and Results}  \label{sec:meas}

\begin{figure}[t!]

\centering
  \includegraphics[width=0.8\columnwidth]{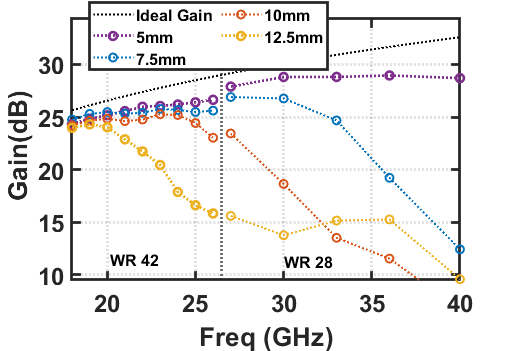}
  \caption{Measured gain. Black dotted line shows 100\% aperture efficiency. Gray dotted line shows WR42 and WR28 measurement bands.}\label{Fig1Label}
\label{Fig:gain}
\vspace*{-0.1in}
\end{figure}

Each Luneburg lenses is measured in an NSI-MI near-field antenna range, shown in Figure\,\ref{Fig:setup}. The measurement involved K-Band and Ka-Band. For each band, a 20\,dBi NSI-MI standard gain horn antenna was measured as the reference antenna, shown in Fig.\,\ref{Fig:setup}(a). The Luneburg lenses were then fed by a 10\,dBi horn in Fig.\,\ref{Fig:setup}(b) and the gain was derived using the comparison method. The measured gain is shown in Figure\,\ref{Fig:gain}. The maximum gain for a 100\% efficient aperture is plotted with the dotted black trace for comparison. 

Below 20\,GHz all four lenses are functional with good performance. However, above 20\,GHz lenses with larger unit-cell sizes show gain reduction at lower frequencies. Table\,\ref{tab:lens} summarizes the maximum operating frequency (defined as the frequency at which gain begins to reduce as frequency is increased) of each prototype. The largest unit-cell (12.5\,mm) prototype operates well up to 20\,GHz. The 10\,mm, 7.5\,mm, and 5\,mm unit-cell prototypes operate up to 25, 33, and $>$40\,GHz, respectively. Note that the 5mm lens maximum frequency is possibly above 40\,GHz, beyond the measuring range available. The third column shows the operation bandwidth of four lenses, and it can be seen that for applications at K Band only, the 10\,mm lens can cover nearly the entire band with one-eighth the number of unit-cells, compared to the 5\,mm lens. The last column gives a ratio of unit-cell size $l_{uc}$ versus the guided wavelength $\lambda_g$, which is derived by $\lambda_g = \lambda_{m} / \sqrt{\epsilon_{r,host}}$ where $\lambda_{m}$ is the wavelength at the maximum stated frequency (column 2 of Tab.\,\ref{tab:lens}). At the maximum operating frequency of each lens, the 7.5, 10, and 12.5\,mm unit-cells are approximately $1.4\lambda_g$ on a side\footnote{The 5\,mm unit-cell measured $>$40\,GHz could exhibit similar size.}. However, the gyroid unit-cell comprises octants of identical, rotated and translated sub-units (see Fig.\,\ref{Fig:concept}(d)). Therefore, the sub-unit is $0.7\lambda_g$ on a side suggesting a design guideline for AM GRIN media. This result is specific to the Luneburg lens profiles investigated in this work. Extending to more general printed lenses is the subject of future investigations.

\begin{table}[t]
\label{tab:lens}
\caption{Lens performance} \label{Table1Label}
\centering
\vspace{-0.1in}
\begin{tabular}{c|c|c|c}
 Unit-cell size $l_{uc}$ & Max. operation freq.  & Bandwidth & $l_{uc}/\lambda_g$  \\
 \hhline {=|=|=|=}
5 mm & $>$ 40 GHz & $>$ 22 GHz& 1.11 \\
 \hline
7.5 mm& 33 GHz & 15 GHz & 1.38 \\
 \hline
10 mm& 25 GHz & 7 GHz & 1.39 \\
  \hline
 12.5 mm& 20 GHz & 2 GHz & 1.39 \\
\end{tabular}

\end{table}

\bibliographystyle{ieeetr}
\bibliography{APSURSI}

\end{document}